\documentclass[preprint,showpacs,preprintnumbers,amsmath,amssymb,nofootinbib]{revtex4}

\usepackage{graphicx}

\begin{document}

\preprint{DF/IST-02.2002}

\title{Gravitational radiation from collisions at the speed of light:
a massless particle falling into a Schwarzschild black hole}

\author{Vitor Cardoso}
\email{vcardoso@fisica.ist.utl.pt}
\author{Jos\'e P. S. Lemos}
\email{lemos@kelvin.ist.utl.pt}
\affiliation{
Centro Multidisciplinar de Astrof\'{\i}sica - CENTRA, 
Departamento de F\'{\i}sica, Instituto Superior T\'ecnico,
Av. Rovisco Pais 1, 1049-001 Lisboa, Portugal
}%

\date{\today}

\begin{abstract}
We compute spectra, waveforms, angular distribution and total
gravitational energy of the gravitiational radiation emitted during the
radial infall of a massless particle into a Schwarzschild black hole.
Our fully relativistic approach shows that (i) less than 50\% of the
total energy radiated to infinity is carried by quadrupole waves, (ii)
the spectra is flat, and (iii) the zero frequency limit agrees
extremely well with a prediction by Smarr.  This process may be looked
at as the limiting case of collisions between massive particles
traveling at nearly the speed of light, by identifying the energy $E$
of the massless particle with $m_0 \gamma$, $m_0$ being the mass of
the test particle and $\gamma$ the Lorentz boost parameter.  
We comment on the implications for the two black
hole collision at nearly the speed of light process, where we obtain a
13.3\% wave generation efficiency, and compare our results with previous 
results by D'Eath and Payne. 
\end{abstract}

\pacs{04.70.Bw, 04.30.Db}

\maketitle

The study of gravitational wave emission by astrophysical objects has
been for the last decades one of the most fascinating topics in General
Relativity. This enthusiasm is of course partly due to the possibility
of detecting gravitational waves by projects such as GEO600 \cite{geo}, 
LIGO \cite{ligo} or VIRGO \cite{virgo}, already
operating.  Since gravity couples very weakly to matter, one needs to
have powerful sources of in order to hope for the detection of the
gravitational waves.  Of the candidate sources, black holes stand out
naturally, as they provide huge warehouses of energy, a fraction of
which may be converted into gravitational waves, by processes such as
collisions between two black holes.  As it often happens, the most
interesting processes are the most difficult to handle, and events such
as black hole-black hole collisions are no exception. An efficient
description of such events requires the use of the full non linear
Einstein's equations, which only begin to be manageable by numerical
methods, and state-of-the-art computing.  In recent years we have
witnessed serious progress in this field \cite{anninos}, and
we are now able to evolve numerically the collision of two black holes,
provided their initial separation is not much larger than a few
Schwarzschild radius. At the same time these numerical results have
been supplemented with results from first and second order perturbation
theory \cite{gleiser}, which simultaneously served as guidance into the
numerical codes. The agreement between the two methods is not only
reassuring, but it is also in fact impressive that a linearization of
Einstein's equations yield such good results (as Smarr \cite{smarr1}
puts it, ``the agreement is so remarkable that something deep must be
at work'').  In connection with this kind of events, the use of
perturbation methods goes back as far as 1970, when Zerilli
\cite{zerilli1} and Davis et al \cite{davis} first computed the
gravitational energy radiated away during the infall from rest at
infinity of a small test particle of mass $m_0$, into a Schwarzschild
black hole with mass $M$.  Later, Ruffini \cite{ruffini} generalized
these results to allow for an initial velocity of the test particle 
(this problem has recently been the subject of further study \cite{lousto}, 
in order to investigate the question of choosing appropriate initial data
for black hole collisions).  Soon after, one began to realize that the
limit $m_0 \rightarrow M$ describing the collision of two black holes
did predict reasonable results, still within perturbation theory,
thereby making perturbation theory an inexpensive tool to study
important phenomena.

In this paper we shall extend the results of Davis et al \cite{davis}
and Ruffini \cite{ruffini} by considering a massless test particle
falling in from infinity through a radial geodesic. This process
describes the collision of an infalling test particle in the limit that
the initial velocity goes to the speed of light, thereby extending the
range of Ruffini's results into larger Lorentz boost parameters 
$\gamma$s.
If then one continues to rely on the agreement between
perturbation theory and the fully numerical outputs, these results
presumably describe the collision of two black holes near the speed of
light, these events have been extensively studied through matching
techniques by D'Eath (a good review can be found in his book
\cite{D'Eath}), and have also been studied in \cite{smarr2}. 
The extension is straightforward, the mathematics involved are quite
standard, but the process has never been studied. Again supposing that
these results hold for the head on collision of two black holes
travelling towards each other at the speed of light, we have a very
simple and tractable problem which can serve as a guide and supplement
the results obtained by Smarr and by D'Eath.  Another strong
motivation for this work comes from the possibility of black hole
formation in TeV-scale gravity \cite{dimopoulos}.  Previous estimates
on how this process develops, in particular the final mass of the
black hole formed by the collision of relativistic particles have
relied heavily upon the computations of D'Eath and Payne \cite{payne}.
A fresher look at the problem is therefore recommended, and a
comparation between our results with results obtained years ago
\cite{D'Eath,smarr2} and with recent results \cite{eardley,solodukhin}
are in order.

Our fully relativistic results show an impressive agreement with
results by Smarr \cite{smarr2} for collisions of massive particles near
the speed of light, namely a flat spectrum, with a zero frequency limit
(ZFL) $(\frac{dE}{d\omega})_{\omega = 0}= 0.4244 m_0^2 \gamma^2$.  We
also show that Smarr underestimated the total energy radiated to
infinity, which we estimate to be $\Delta E=0.26 m_0^2 \gamma^2/M$,
with $M$ the black hole mass.  The quadrupole part of the perturbation
carries less than 50\% of this energy.  When applied to the head on
collision of two black holes moving at the speed of light, we obtain an
efficiency for gravitational wave generation of 13\%, quite close to
D'Eath and Payne's result of 16\% \cite{D'Eath,payne}.

Since the mathematical formalism for this problem has been thoroughly
exploited over the years, we will just outline the procedure. Treating the
massless particle as a perturbation, we write the metric functions for
this spacetime, black hole + infalling particle,  as
\begin{equation}
g_{ab}(x^\nu)= g^{(0)}_{ab}(x^\nu)+h_{ab}(x^\nu)\,,
\label{perturbation}
\end{equation}
where the metric $g^{(0)}_{ab}(x^\nu)$ is the background metric,
(given by some known solution of Einstein's equations), which we now specialize
to the Schwarzschild metric
\begin{equation}
ds^{2}=-f(r)dt^{2}+\frac{dr^{2}}{f(r)}+
r^{2}(d\theta^{2}+\sin^2\theta d\phi^{2})\,,
\label{lineelement}
\end{equation}
where $f(r)=1-2M/r$.
Also, $ h_{ab}(x^\nu)$ is a small perturbation, induced by the
massless test particle, which is described by the stress energy tensor
\begin{equation}
T^{\mu \nu}= -\frac{p_0}{(-g)^{1/2}} \int d \lambda 
\delta^4 (x-z(\lambda))\dot{z}^{\mu} \dot{z}^{\nu}\,. 
\label{stresstensor}
\end{equation}
Here, $z^{\nu}$ is the trajectory of the particle along the world-line,
parametrized by an affine parameter $\lambda$ (the proper time in the case 
of a massive particle), 
and $p_0$ is the momentum of the particle. 
To proceed, we decompose Einstein's equations $G_{ab}=8\pi T_{ab}$
in tensorial spherical harmonics and
specialize to the Regge-Wheeler \cite{regge} gauge. 
For our case, in which the particle falls
straight in, only even parity perturbations survive. Finally,
following Zerilli's \cite{zerilli2} prescription, we arrive at a
wavefunction (a function of the time $t$ and radial $r$ coordinates only)
whose evolution can be followed by the wave equation
\begin{equation}
\frac{\partial^{2} {\bf \tilde Z}(\omega,r)}{\partial r_*^{2}} +
\left\lbrack\omega^2-V(r)\right\rbrack{\bf \tilde Z}(\omega,r)=(1-2M/r)S \,,
\label{waveequation}
\end{equation}
Here, the $l-$dependent potential $V$ is given by
\begin{equation}
V(r)=\frac{f(r)\left[2\sigma^2(\sigma+1)r^3+6\sigma^2 r^2M +18 
\sigma rM^2 +18M^3\right]}
{r^3(3M+\sigma r)^2}\,,
\label{potentialmaxwell}
\end{equation}
where $\sigma=\frac{(l-1)(l+2)}{2}$ and the tortoise 
coordinate $r_*$ is defined as
$\frac{\partial r}{\partial r_*}= f(r)\,$. 
The passage from the time variable $t$ to the frequency $\omega$ has been achieved
through a Fourier transform, 
${\bf \tilde  Z}(\omega,r)=
\frac{1}{(2\pi)^{1/2}}\int_{-\infty}^{\infty}e^{i \omega t}Z(t,r)dt$.
The source $S$ depends entirely on the stress energy tensor of the 
particle and on whether or not it is massive.
The difference between massive and massless particles
lies on the geodesics they follow.
The radial geodesics for massive particles are:
\begin{equation}
\frac{dT}{dr}=-\frac{E}{f(r)(E^2-1+2M/r)^{1/2}}\,;
\frac{dt}{d\tau} =\frac{E}{f(r)}\,,
\label{massgeo}
\end{equation}
where $E$ is a conserved energy parameter: For example, if the particle
has velocity $v_{\infty}$ at infinity then
$E=\frac{1}{(1-v_{\infty}^2)^{1/2}}\equiv\gamma$.  On the other hand,
the radial geodesics for massless particles are described by
\begin{equation}
\frac{dT}{dr}=-\frac{1}{f(r)}\,\,;\, 
\frac{dt}{d\tau} =\frac{\epsilon_0}{f(r)}\,,
\label{masslessgeo}
\end{equation}
where again $\epsilon_0$ is a conserved energy parameter, which in relativistic
units is simply $p_0$. We shall however keep $\epsilon_0$ for future use, to
see more directly the connection between massless particles and massive ones
traveling close to the speed of light.
One can see that, on putting $p_0 \rightarrow m_0$,  
$\epsilon_0 \rightarrow \gamma$ and  $\gamma \rightarrow \infty$, the radial
null geodesics reduce to radial timelike geodesics, so that all the results
we shall obtain in this paper can be carried over to the case of ultrarelativistic
(massive) test particles falling into a Schwarzschild black hole.
For massless particles the source term $S$ is 
\begin{equation}
S=\frac{4ip_0 e^{-i\omega r_*}\epsilon_0 (4l+2)^{1/2}\sigma }
{w(3M+ \sigma r)^2}\,.
\label{source}
\end{equation}
To get the energy spectra, we use
\begin{equation}
\frac{dE}{d\omega}=\frac{1}{32 \pi}\frac{(l+2)!}{(l-2)!} \omega^2 |{\bf \tilde Z}(\omega,r)|^2\,,
\label{spectra}
\end{equation}
and to reconstructe the wavefunction $Z(t,r)$ one uses the inverse Fourier transform
\begin{equation}
Z(t,r)=
\frac{1}{(2\pi)^{1/2}}\int_{-\infty}^{\infty}e^{-i \omega t}{\bf \tilde Z}(\omega,r)d\omega.
\label{invfourier}
\end{equation}
Now we have to find ${\bf \tilde Z}(\omega,r)$ from the differential
equation (\ref{waveequation}).  This is accomplished by a Green's
function technique. Imposing the usual boundary conditions, i.e., only
ingoing waves at the horizon and outgoing waves at infinity, we get
that, near infinity,
\begin{equation}
{\bf \tilde Z}=\frac{1}{W}\int_{r_+}^{\infty}z_L S dr.
\label{solution}
\end{equation}
Here, $z_L$ is a homogeneous solution of (\ref{waveequation}) which
asymptotically behaves as
\begin{eqnarray}
z_L \sim e^{-i\omega r_*}\,,r_* \rightarrow -\infty \\
z_L \sim B(\omega)e^{i\omega r_*}+C(\omega)e^{-i\omega r_*}\,,r_* \rightarrow +\infty. 
\label{behavior1}
\end{eqnarray}
$W$ is the wronskian of the homogeneous solutions of (\ref{waveequation}). 
These solutios are, 
$z_L$ which has just been defined, and $z_R$ which behaves as $z_R \sim
e^{i\omega r_*}\,,r_* \rightarrow +\infty $. From this follows that
$W=2i\omega C(\omega)$.  We find $C(\omega)$ by solving
(\ref{waveequation}) with the right hand side set to zero, and with the
starting condition $z_L= e^{-i\omega r_*}$ imposed at a large negative
value of $r_*$. For computational purposes good accuracy is hard to
achieve with the form (\ref{behavior1}), so we used an asymptotic
solution one order higher in $1/(\omega r)$:
\begin{eqnarray}
&z_L = B(\omega)(1+\frac{i(\sigma+1)}{\omega r})e^{i\omega r_*}+
\nonumber\\ &
+C(\omega)(1-\frac{i(\sigma+1)}{\omega r})e^{-i\omega r_*}\,,r_* \rightarrow +\infty\,.&
\label{behavior2}
\end{eqnarray}
In the numerical work, we chose to adopt $r$ as the independent
variable, thereby avoiding the numerical inversion of $r_*(r)$.
A fourth order Runge-Kutta routine started the integration
of $z_L$ near the horizon, at $r=r_i=2M+2M\epsilon$, with tipically 
$\epsilon=10^{-5}$. It then integrated out to large values
of $r$, where one matches $z_L$ extracted numerically with the
asymptotic solution (\ref{behavior2}), in order to find $C(\omega)$.
To find $Z(t,r)$ the integral in (\ref{invfourier}) is
done by Simpson's rule. For both routines Richardson
extrapolation is used.
The results for the wavefunction $Z(t,r)$ as a function of the
retarded time $u\equiv t-r_*$ are shown in Figure 1, for the
three lowest radiatable multipoles, $l=2, 3$ and $4$.  
\begin{figure}
\includegraphics{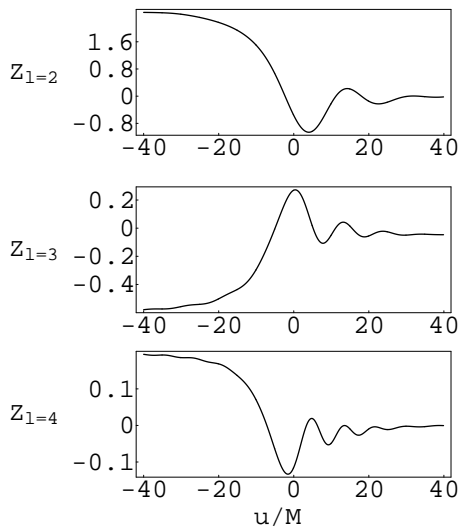}
\caption{\label{waveformgraph}
 Waveforms for the three lowest
radiatable multipoles, for a massless particle falling from 
infinity into a Schwarzschild black hole. Here, the wavefunction 
$Z$ is measured in units of $\epsilon_0p_0$.
}
\end{figure}
As expected from the work of Ruffini \cite{ruffini} and Ferrari and
Ruffini \cite{ferrari}, the wavefunction is not zero at very early
times, reflecting the fact that the particle begins to fall with non
zero velocity. At late times, the $l=2$ (for example) signal is 
dominated by quasinormal ringing with 
frequency $\omega \sim 0.35/M$, the lowest quasinormal
frequency for this spacetime \cite{chandra}.  The energy
spectra is shown in Figure 2, for the four lowest radiatable
multipoles. First, as expected from Smarr's work \cite{smarr2}, the
spectra is flat, up to a certain critical value of the frequency, after
which it rapidly decreases to zero. This ($l$-dependent) critical
frequency is well approximated, for each $l$-pole, by the fundamental
quasinormal frequency. In Table 1, we list the zero frequency limit
(ZFL) for the first ten lowest radiatable multipoles.
\begin{table}
\caption{\label{tab:zfl}  The zero frequency limit (ZFL) 
for the ten lowest radiatable multipoles.}
\begin{ruledtabular}
\begin{tabular}{llll}  \hline
$l$ & ZFL($\times\frac{1}{\epsilon_0^2p_0^2}$)&$l$& ZFL($\times \frac{1}{\epsilon_0^2p_0^2}$)\\ \hline
2    &  0.265 &  7 &  0.0068 \\ \hline 
3    &  0.075 &   8 &  0.0043  \\ \hline 
4    &  0.032 &   9 &  0.003 \\ \hline 
5    &  0.0166 &   10 &  0.0023  \\ \hline 
6    &  0.01  &   11 &  0.0017 \\ \hline 
\end{tabular}
\end{ruledtabular}
\end{table}

For high
values of the angular quantum number $l$, a good fit to our numerical
data is
\begin{equation}
\left(\frac{dE_l}{d\omega}\right)_{\omega = 0} = 
\frac{2.25}{l^3} \epsilon_0^2 p_0^2
\label{fit}
\end{equation}
We therefore estimate the zero ZFL as
\begin{eqnarray}
&\left(\frac{dE_l}{d\omega}\right)_{\omega = 0}=
\left[\sum_{l=2}^{l=11}
\left(\frac{dE_l}{d\omega}\right)_{\omega = 0}\right] +
\frac12 \frac{2.25}{12^2} \epsilon_0^2p_0^2
\nonumber\\ &
=0.4244 \epsilon_0^2p_0^2\,.&
\label{ZFL}
\end{eqnarray}
To calculate the total energy radiated to infinity, we proceed as
follows: as we said, the spectra goes as $2.25/l^3$ as long as $\omega
< \omega_{lQN}$, where $\omega_{lQN}$ is the lowest quasinormal
frequency for that $l$-pole.  For $\omega > \omega_{lQN}$, $dE/d\omega
\sim 0$ (In fact, our numerical data shows that $dE/d\omega \sim
e^{-27\alpha\omega M}$, with $\alpha$ a factor of order unity, for
$\omega > \omega_{lQN}$).  Now, from the work of Ferrari and Mashhoon
\cite{ferrari2} and Schutz and Will \cite{schutz}, one knows that for
large $l$, $\omega_{lQN} \sim \frac{l+1/2}{3^{3/2}M}$.  Therefore, for
large $l$ the energy radiated to infinity in each multipole is
\begin{equation}
\Delta E_l=\frac{2.25(l+1/2)}{3^{3/2}l^3}\frac{\epsilon_0^2p_0^2}{M}\,, 
\label{totalEl}
\end{equation}
and an estimate to the total energy radiated is then
\begin{equation}
\Delta E=\sum_l \Delta E_l = 0.262\frac{\epsilon_0^2p_0^2}{M}
\label{totalE}
\end{equation}
\begin{figure}
\includegraphics{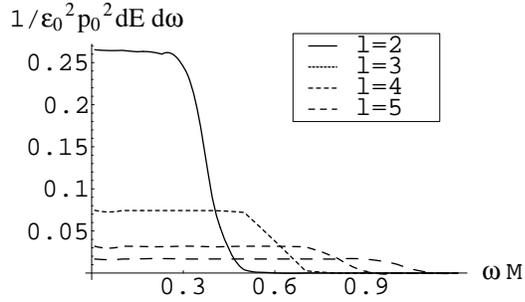}
\caption{\label{spectragraph}
The energy spectra for the four
lowest radiatable multipoles, for a massless particle falling
from infinity into a Schwarzschild black hole. 
}
\end{figure}
Let us now make the bridge between these results and previous results
on collisions between massive particles at nearly the speed of light
\cite{ruffini,D'Eath,smarr2}. As we mentioned, putting
$p_0 \rightarrow m_0$ and $\epsilon_0 \rightarrow \gamma$ does the trick.
So for ultrarelativistic test particles with mass $m_0$ falling into a 
Schwarzschild black hole, one should have 
$(dE/d\omega)_{\omega=0}= 0.4244 m_0^2 \gamma^2$ and 
$\Delta E=  0.262\frac{ m_0^2 \gamma^2}{M}$.
Smarr \cite{smarr2} obtains
\begin{eqnarray}
&\left(\frac{dE}{d\omega}\right)^{\rm Smarr}_{\omega=0}=
\frac{8}{6\pi}m_0^2 \gamma^2\sim 0.4244 m_0^2 \gamma^2\,,
\nonumber\\ &
\Delta E_{\rm Smarr}= 0.2 \frac{m_0^2 \gamma^2}{M}\,.&
\label{smarrresult}
\end{eqnarray}
So Smarr's result for the ZFL is in excellent agreement with ours,
while his result for the total energy radiated is seen to be an
underestimate. As we know from the work of Davis et al \cite{davis}
for a particle falling from infinity with $v_{\infty}=0$ most of the
radiation is carried by the $l=2$ mode.  Not so here, in fact in our
case less than 50\% is carried in the quadrupole mode (we obtain
$\Delta E_{l=2} = 0.1\frac{\epsilon_0^2p_0^2}{M}$, $\Delta E_{l=3} =
0.0447\frac{\epsilon_0^2p_0^2}{M}$ ).
This is reflected in the angular distribution of the radiated energy
(power per solid angle)
\begin{equation}
\frac{dE}{d\Omega}=\Delta E_l \frac{(l-2)!}{(l+2)!}
\left[2 \frac{\partial ^2}{\partial \theta ^2} Y_{l0} +
l(l+1)Y_{l0}\right]^2,
\label{powerangle}
\end{equation}
which we plot in Figure 3. Compare with figure 5 of \cite{smarr2}.

\begin{figure}
\includegraphics{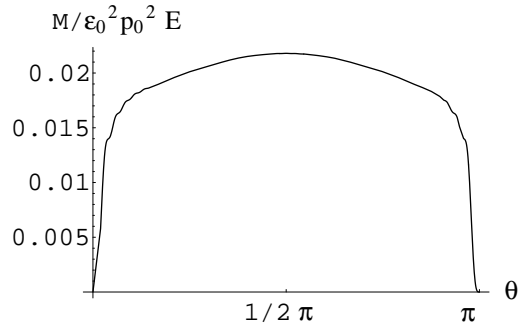}
\caption{\label{angulardist}
The energy radiated per solid angle as
a function of $\theta$. 
}
\end{figure}
If, as Smarr, we continue to assume that something deep is at work, and
that these results can be carried over to the case of two equal mass
black holes flying towards each other at (close to) the speed of light,
we obtain a wave generation efficiency of 13 \%. This is in close
agreement with results by D'Eath and Payne \cite{D'Eath,payne}, who
obtain a 16\% efficiency (Smarr's results cannot be trusted in this
regime, as shown by Payne \cite{payne2}). Now, D'Eath and Payne's
results were achieved by cutting an infinite series for the news
function at the second term, so one has to take those results with some
care.  However, the agreement we find between ours and their results
lead us to believe that once again perturbation theory has a much more
wider realm of validity.  To our knowledge, this is the first
alternative to D'Eath and Payne's method of computing the energy
release in such events.


This work was partially funded by Funda\c c\~ao para a
Ci\^encia e Tecnologia (FCT) through project PESO/PRO/2000/4014. V.C.
also acknowledges finantial support from FCT through PRAXIS XXI
programme.  J. P. S. L. thanks Observat\'orio Nacional do Rio de
Janeiro for hospitality.



\end{document}